\documentclass[journal]{IEEEtran}

\usepackage{pgfplots}
\usepackage[bookmarks,colorlinks]{hyperref}
\usepackage[linesnumbered,ruled,lined]{algorithm2e}
\usepackage{enumitem}
\usetikzlibrary{shapes.multipart,intersections}
\usepackage{cite}
\usepackage{amsmath,amssymb,amsfonts,amsthm,steinmetz}
\usepackage{mathrsfs}  
\usepackage{textcomp}
\usepackage{acronym}
\usepackage{xcolor}
\usepackage{upgreek,xspace}
\usepackage{array}
\usepackage{tikz}
\usetikzlibrary{calc}
\makeatletter
\newcommand{\gettikzxy}[3]{%
  \tikz@scan@one@point\pgfutil@firstofone#1\relax
  \edef#2{\the\pgf@x}%
  \edef#3{\the\pgf@y}%
}
\makeatother

\usepackage[draft]{todonotes}

\usepackage{esvect}

\usepackage{times}
\usepackage{bm}
\usepackage{stmaryrd}
\usepackage{babel}
\usepackage{graphics, graphicx}
\usepackage{gensymb}
\usepackage{cite}
\usepackage{enumitem}
\usepackage{url}

\begin{document}

\title{Wireless Multi-Port Sensing: Virtual-VNA-Enabled De-Embedding of an Over-the-Air Fixture}

\author{Philipp~del~Hougne,~\IEEEmembership{Member,~IEEE}
\thanks{This work was supported in part by the ANR France 2030 program (project ANR-22-PEFT-0005), the ANR PRCI program (project ANR-22-CE93-0010), the European Union's European Regional Development Fund, and the French Region of Brittany and Rennes M\'etropole through the contrats de plan \'Etat-R\'egion program (projects ``SOPHIE/STIC \& Ondes'' and ``CyMoCoD'').}
\thanks{
P.~del~Hougne is with Univ Rennes, CNRS, IETR - UMR 6164, F-35000, Rennes, France (e-mail: philipp.del-hougne@univ-rennes.fr).
}
}

\maketitle

\begin{abstract}
We develop a multi-port-backscatter-modulation technique to determine, over the air (OTA), the scattering parameters of a linear, passive, time-invariant multi-port device under test (DUT).
A set of ``not-directly-accessible'' (NDA) antennas can be switched between being terminated by the DUT or by a specific, known, tunable load network. Waves can be radiated and captured via a distinct set of ``accessible'' antennas that couple OTA to the NDA antennas. \textit{First}, we characterize the OTA fixture between the accessible antennas' ports and the DUT's ports. We achieve this based on our recently introduced ``Virtual VNA'' technique; specifically, we connect the NDA antennas to the tunable load network and measure the scattering at the accessible antennas' ports for various configurations of the tunable load network. \textit{Second}, we connect the NDA antennas to the DUT and measure the scattering at the accessible antennas' ports. \textit{Third}, we de-embed the OTA fixture to retrieve the DUT's scattering parameters. 
We experimentally validate our technique at 2.45~GHz for 1-port DUTs and 5-port DUTs, considering a rich-scattering OTA fixture inside a reverberation chamber. We systematically study the influence of the number of accessible antennas and various conceivable simplifications in terms of the system model as well as the properties of the tunable load network. 
Our wireless multi-port sensing technique can find applications in areas like RFID and wireless bioelectronics.
\end{abstract}


\section{Introduction}
\label{sec_introduction}

Communications and sensing techniques underpinned by antenna backscatter modulation have been explored for decades in contexts spanning from the Great Seal Bug in the 1940s~\cite{brooker2013lev} to the nowadays ubiquitous radio-frequency identification (RFID) technology. More recently, the tailoring of wireless channels with reconfigurable intelligent surfaces (RISs) constitutes a further prominent example. 

Wireless sensing can be classified according to how the sensor information is encoded into wireless signals~\cite{Marrocco}. If the sensor information is encoded \textit{digitally} (e.g., by modulating the ID of a tag), then the role of antenna backscatter modulation is purely one of facilitating digital wireless communications. If, however, the sensing mechanism results in an \textit{analog} modulation of an antenna's reflection coefficient or load impedance, there is no digital modulation. In this case, the sensing accuracy directly depends on the accuracy with which the antenna's reflection coefficient or load impedance can be estimated remotely. The influence of the wireless propagation environment (WPE) can be substantial. In particular, unless the WPE is free space, a suitable form of in situ parameter estimation to characterize the WPE may be required before the sought-after sensor information can be inferred. 

The complexity of extracting sought-after sensor information based on analog modulation increases further in the case of multi-port backscatter modulation, also referred to as RFID grids~\cite{Marrocco_RFID_GRID,Marrocco_RFID_GRID2}. This case arises either when multiple backscatter-modulated single-port antennas are placed in close proximity~\cite{marrocco2008multiport,mughal2023statistical,nanni2024stackedRFID,barbot2025differential} (e.g., densely distributed sensor networks like stacked RFID sensors) or in scenarios involving a single multi-port backscatter-modulated system~\cite{caizzone2011multi}. The key challenge in these multi-port backscatter-modulation cases lies in the mutual coupling between the ports~\cite{Marrocco_RFID_GRID,Marrocco_RFID_GRID2}. 

Wireless sensing based on analog modulation can hence be classified according to whether, \textit{first}, the sensor information is encoded in the antenna's reflection coefficient or the antenna's load impedance, and, \textit{second}, whether a single port or multiple ports are backscatter-modulated. We now briefly review the current state-of-the-art in each of these four cases in turn.

\textit{Remote single-port reflection-coefficient sensing.} For the case in which the WPE is free space, the estimation of the reflection coefficient of a single-port antenna by modulating its load impedance has been tackled in~\cite{garbacz1964determination,mayhan1994technique,pursula2008backscattering,bories2010small,van2020verification,sahin2021noncontact,kruglov2023contactless}. Three distinct, known loads of the antenna are required. If the WPE is not free space, the antenna's reflection coefficient generally depends on the WPE specifics. This dependence might prevent a direct extraction of the sought-after sensor information from the antenna's reflection coefficient. Recently,~\cite{del2024virtual} presented a general treatment not assuming a free-space WPE and making optimal use of multiple remote probing antennas, either via a matrix-valued closed-form approach or via a gradient-descent approach. The latter was also shown to be compatible with purely non-coherent detection~\cite{del2024virtual}, which greatly alleviates the hardware requirements.

\textit{Remote multi-port reflection-matrix sensing.}\footnote{We use the terms ``reflection matrix'' and ``scattering matrix'' interchangeably throughout this paper. Here, we use the former to highlight that it is an extension of the previously discussed reflection-coefficient sensing.} For the case in which the WPE is free space, remotely estimating the $2\times 2$ reflection matrix of a two-port antenna system by modulating the individual load impedances terminating its ports has been tackled in~\cite{wiesbeck1998wide,monsalve2013multiport,shilinkov2024antenna}. As in the single-port case, a non-trivial WPE generally impacts the sought-after reflection matrix and might thereby hamper the sensing. In any case, the estimation of multi-port reflection matrices in the general case of arbitrarily complex WPEs and making optimal use of multiple remote probe antennas was recently presented in~\cite{del2024virtual}. Specifically, as in the single-port case, the techniques from~\cite{del2024virtual} exist as matrix-valued closed-form or gradient-descent variants, and the latter is compatible with purely non-coherent detection. Because~\cite{wiesbeck1998wide,monsalve2013multiport,shilinkov2024antenna,del2024virtual} were limited to using three distinct, known, individual loads at each antenna port, inevitable sign ambiguities about off-diagonal reflection-matrix entries remained. In some cases, these can be lifted with a priori knowledge~\cite{shilinkov2024antenna}. General techniques for unambiguously estimating the full reflection matrix without any reliance on a priori knowledge require in addition the possibility of terminating the antenna ports with known, coupled loads~\cite{denicke2012application,del2024virtual2p0,tapie2025scalable}. In its simplest form, a coupled load is a 2-port load network (2PLN) that can be embodied by a simple transmission line.

Beyond this antenna-related context, remotely estimating a reflection coefficient or reflection matrix via backscatter modulation can be seen in the wider context of metrology. Given the limited (typically low) number of available vector network analyzer (VNA) ports, measuring the network parameters of a many-port device under test (DUT) traditionally requires countless manual reconnections between VNA and DUT, terminating each time the unconnected DUT ports with matched loads. To reduce the number of required reconnections, methods under names such as  ``untermination'' and ``port-reduction'' were explored~\cite{bauer1974embedding,davidovitz1995reconstruction,lu2000port,lu2003multiport,pfeiffer2005recursive,pfeiffer2005equivalent,pfeiffer2005characterization}. These methods rely on subsequently terminating unconnected DUT ports with distinct, known, individual loads. The set of DUT ports connected to the VNA varies over the course of the measurement procedure.

In contrast, the ``Virtual VNA'' technique~\cite{del2024virtual2p0} operates with a fixed set of ``accessible'' DUT ports (connected to the VNA) and a fixed set of ``not-directly-accessible'' (NDA) DUT ports (connected to a tunable load network). Thereby, no manual reconnections are required. The tunable load network can terminate each DUT NDA port with three distinct individual loads, connect neighboring DUT ports via coupled loads, and connect at least one accessible DUT port and one NDA DUT port via a coupled load. The characteristics of this tunable load network can be chosen arbitrarily but must be known. Each port of the tunable load network acts effectively like an additional ``virtual'' VNA port, hence the term ``Virtual VNA''.
The Virtual VNA not only yields the reflection matrix of the DUT's NDA ports, but also the transmission coefficients between the DUT's accessible and NDA ports, as well as the reflection matrix at the DUT's accessible ports. Moreover, it has been generalized to non-reciprocal DUTs based on the same tunable load network~\cite{del2025virtual3p0,del2025virtual3p1} which can be realized in a scalable manner on printed-circuit board~\cite{tapie2025scalable}. To map the Virtual VNA technique to the wireless sensing problem, one simply identifies the entire WPE (whatever its complexity, and including all structural elements of all antennas) as the DUT. Then, the ports of the remote probe antennas are the accessible DUT ports and the ports of the NDA antennas are the NDA DUT ports. The block of the DUT's scattering matrix exclusively associated with the NDA ports is the sought-after reflection matrix.

Having so far discussed wireless sensing of an antenna's reflection coefficient or an antenna system's reflection matrix, we now survey wireless sensing of an antenna's load impedance and an antenna system's load impedance matrix.

\textit{Remote single-port load-impedance sensing.}
In~\cite{chen2012wireless}, an unknown load impedance was estimated assuming (i) a free-space WPE, (ii) a known antenna design for which the reflection coefficient has been numerically evaluated, and (iii) the availability of two known individual loads.
In~\cite{bjorninen2011wireless,akbar2015rfid}, an unknown load impedance was estimated based on measurements in free space with three (effectively) different antennas for which the reflection coefficients have been numerically evaluated. 
In~\cite{skrobacz2024new}, the number of required known antennas was reduced to two by assuming the real part of the sought-after load impedance is known.
In~\cite{chen2012coupling,vena2024backscatter},  an unknown load impedance was estimated solely based on two known individual loads; as we explain in Sec.~\ref{subsec_simplified_system_model} below, these methods thus tacitly assume that the antenna's reflection coefficient vanishes. 
Altogether, a general method not relying on assumptions of a free-space WPE, the availability of one or multiple perfectly known antenna(s), or the availability of a perfectly matched antenna, is missing to date.

\textit{Remote multi-port load-impedance-matrix sensing.} A technique for estimating a vector whose entries are proportional to the magnitudes of the diagonal entries of the admittance representation of the sought-after load impedance matrix was presented in~\cite{Marrocco_RFID_GRID,Marrocco_RFID_GRID2}. This method relied on sequentially flipping the state of one load away from its reference state while maintaining all other loads in their reference states. However, unambiguously estimating the full load impedance (or admittance) matrix remains illusive to date. Indeed, the method from~\cite{Marrocco_RFID_GRID,Marrocco_RFID_GRID2} neither identifies the unknown proportionality constant(s) concerning the diagonal admittance matrix entries, nor does it identify the off-diagonal admittance matrix entries at all.

In this paper, we unambiguously estimate an antenna system's load impedance matrix OTA and without any assumptions about the WPE (other than linearity, passivity, time-invariance, and reciprocity) or the antennas (other than that their ports are lumped and monomodal). We refer to the load system whose impedance matrix we seek as DUT, which should not be confused with the meaning of DUT in the Virtual VNA context. Here, each DUT port is connected to a distinct NDA antenna, i.e., no DUT port is directly connected to a VNA. The NDA antennas couple OTA to accessible, remote probe antennas via which we can inject and/or receive waves.
Our requirements are the availability of (i) three distinct, known, individual loads that can terminate each NDA antenna, and, (ii) coupled loads that can terminate pairs of NDA antennas. Requirement (i) can be relaxed to two distinct, known, individual loads for multi-port DUTs. Requirement (ii) is waived when the transmission between the DUT's ports is known to be zero (e.g., whenever the DUT is an ensembles of single-port loads), or if only its magnitude is of interest. If the DUT has a single port, requirement (ii) is irrelevant. 

Our \textit{first} important insight consists in viewing the WPE (including environmental scattering and structural scattering from the antennas) as an ``OTA fixture'' between our VNA and the DUT. The benefit of this view lies in its generality: it applies irrespective of any details of environmental scattering in the WPE and irrespective of any details of the antenna design -- as long as our aforementioned very general assumptions remain valid.
To characterize the DUT, the OTA fixture must be de-embedded, which requires knowledge of its characteristics. Our \textit{second} important insight is that a sufficiently unambiguous characterization of the OTA fixture is possible based on the recently established Virtual VNA technique~\cite{del2024virtual2p0}. By ``sufficiently unambiguous'' we mean that we tolerate one sign ambiguity that does \textit{not} propagate into our estimate of the DUT characteristics and that arises from the absence of the possibility to connect one accessible antenna and one NDA antenna via a coupled load.

Our contributions are summarized as follows. 
\textit{First}, we present the first method capable of remotely and unambiguously estimating a full multi-port load impedance matrix, without any assumptions about specifics of the WPE or the utilized antennas. Clearly, our method also applies in the simpler single-port case in which such a general method is also missing to date.
\textit{Second}, we experimentally validate our method in a rich-scattering WPE, both for 1-port DUTs and 5-port DUTs.
\textit{Third}, we systematically examine the influence of key parameters (number of accessible antennas, separation or not of the accessible antennas into transmitting and receiving ones, number of measurements with the tunable load network).
\textit{Fourth}, we examine the effect of conceivable simplifications of the tunable load network requirements and/or the physics-compliant system model. This benchmarking reveals that the Virtual VNA requirements for the tunable load network can be relaxed in cases involving at least two NDA ports.

The remainder of this paper is organized as follows. 
In Sec.~\ref{sec_problem_statement}, we formalize our problem statement in terms of multi-port network theory.
In Sec.~\ref{sec_method}, we present our wireless multi-port sensing method, as well as the considered simplification benchmarks.
In Sec.~\ref{sec_experimental_validation}, we present our experimental validation.
In Sec.~\ref{sec_conclusions}, we briefly conclude and outline future directions.

Throughout this paper, we use the scattering parameter representation, which is fully equivalent to impedance or admittance representations of network characteristics.

\section{Problem Statement}
\label{sec_problem_statement}

Following the taxonomy outlined in Sec.~\ref{sec_introduction}, we tackle the problem of \textit{remote multi-port load-impedance-matrix sensing}. 
We thus assume that the unknown load system (referred to as DUT in the following) is \textit{not} itself an antenna array, or, more specifically, not an antenna array radiating into the WPE that constitutes our OTA fixture. If the latter were the case, the wireless sensing problem could be reframed as remote multi-port reflection-matrix sensing for which prior work already introduced the Virtual VNA technique. We further discuss our problem statement's restriction for radiative DUTs at the end of this section.

\begin{figure}
    \centering
    \includegraphics[width=\columnwidth]{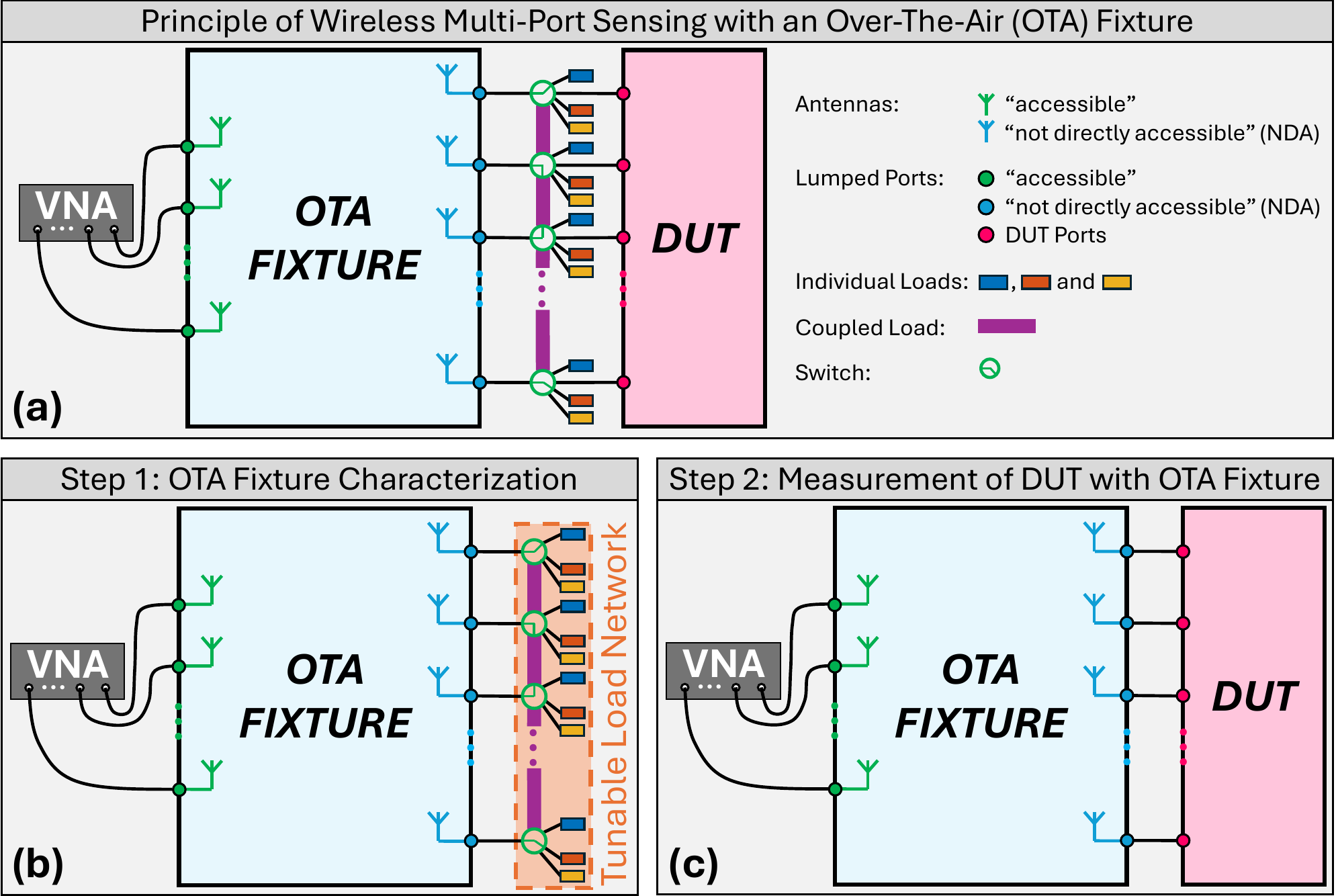}
    \caption{(a) Principle of wireless multi-port sensing with an OTA fixture. The DUT to be characterized can be connected to a set of NDA antennas; alternatively, these NDA antennas can be terminated by a known, tunable load network. The tunable load network can terminate NDA antennas with three distinct individual loads or pairs of neighboring NDA antennas via coupled loads. 
    The NDA antennas couple OTA to a set of accessible antennas connected to a VNA. 
    (b) In Step 1, the OTA fixture is characterized using realizations of the known, tunable load network based on the Virtual VNA technique. (c) In Step 2, the unknown DUT is measured with the OTA fixture. In Step 3, the OTA fixture is de-embedded from the measurement taken in Step 2 to recover the DUT's scattering parameters. }
    \label{Fig1}
\end{figure}

We consider a DUT with $N_\mathrm{S}$ lumped, monomodal ports. We assume that the DUT is linear, passive, and time-invariant.
Our goal is to \textit{wirelessly} determine the DUT's scattering matrix $\mathbf{S}^\mathrm{D}\in\mathbb{C}^{N_\mathrm{S}\times N_\mathrm{S}}$, i.e., without connecting any of the DUT's ports directly to a VNA. Instead, we connect each DUT port to a distinct antenna. This set of $N_\mathrm{S}$ antennas is NDA; we formally define NDA as meaning that we cannot inject or receive waves via the ports of these antennas. However, we can switch the termination of these NDA antennas. Specifically, as shown in Fig.~\ref{Fig1}a, the NDA antennas can be terminated either by the DUT or by a specific tunable load network.
The latter satisfies the following requirements of the ``Virtual VNA Extension Kit''~\cite{del2024virtual2p0}: (i) each NDA antenna port can be connected to one of three distinct individual loads; (ii) pairs of neighboring NDA antenna ports can alternatively be connected via 2PLNs; (iii) the reflection coefficients of the individual loads and the $2\times 2$ scattering matrices of the 2PLNs are known. The three individual loads can be chosen arbitrarily (as long as they are mutually distinct) and may differ for each NDA antenna. The 2PLNs can be chosen arbitrarily (as long as their transmission coefficients do not vanish) and may differ from each other.
Besides the NDA antennas, there are $N_\mathrm{A}$ ``accessible'' antennas that are located remotely with respect to the DUT. By ``accessible'' we mean that we can inject and/or receive waves via these antennas. These accessible antennas are coupled OTA to the NDA antennas. We make no assumption about the WPE other than that it is linear, passive, time-invariant, and reciprocal.

Because of the OTA coupling between the accessible antennas and the NDA antennas, the scattering that we can measure at the accessible antennas' ports depends on the termination of the NDA antennas' ports. We assume that all accessible antennas and all NDA antennas have lumped, monomodal ports. Then, the three distinct scattering matrices involved in our problem statement can be defined as follows:
\begin{itemize}
    \item $\mathbf{S}\in\mathbb{C}^{N_\mathrm{A}\times N_\mathrm{A}}$ is the scattering matrix that can be measured at the ports of the $N_\mathrm{A}$ accessible antennas. If the latter are partitioned into transmitting and receiving antennas, we measure a transmission matrix $\mathbf{H}=\mathbf{S}_\mathcal{RT}$, where the subscript $\mathcal{RT}$ denotes selecting the block of $\mathbf{S}$ whose row [column] indices are in the set $\mathcal{R}$ [$\mathcal{T}$] which comprises the indices of the  ports associated with receiving [transmitting] accessible antennas. 
    \item $\mathbf{S}^\mathrm{L}\in\mathbb{C}^{N_\mathrm{S} \times N_\mathrm{S}}$ is the scattering matrix characterizing the NDA antennas' termination. Depending on how we configure the NDA antennas' termination, $\mathbf{S}^\mathrm{L}$ is either the scattering matrix of the tunable load network or the DUT (both of which have $N_\mathrm{S}$ ports). $\mathbf{S}^\mathrm{L}$ is generally \textit{not} diagonal. If $\mathbf{S}^\mathrm{L}$ is realized by the tunable load network, it is known; if $\mathbf{S}^\mathrm{L}$ equals $\mathbf{S}^\mathrm{D}$, it is unknown and determining it is our ultimate goal.
    \item $\mathbf{S}^\mathrm{F}\in\mathbb{C}^{N \times N}$ is the scattering matrix of the ``OTA fixture'' connecting the ports of the accessible antennas and the ports of the NDA antennas, where $N=N_\mathrm{A}+N_\mathrm{S}$. $\mathbf{S}^\mathrm{F}$ captures both environmental and structural scattering within the WPE. In other words, the structure of the accessible and NDA antennas is a part of the WPE. $\mathbf{S}^\mathrm{F}$ is fixed and unknown; determining $\mathbf{S}^\mathrm{F}$ is our intermediate goal to subsequently de-embed the OTA fixture.
\end{itemize}
The relation between $\mathbf{S}$, $\mathbf{S}^\mathrm{F}$ and $\mathbf{S}^\mathrm{L}$ is known from multi-port network theory~\cite{anderson_cascade_1966,ha1981solid}:
\begin{equation}
    \mathbf{S} = \mathbf{S}^\mathrm{F}_\mathcal{AA} + \mathbf{S}^\mathrm{F}_\mathcal{AS} \left( \left(\mathbf{S}^\mathrm{L}\right)^{-1} - \mathbf{S}^\mathrm{F}_\mathcal{SS}\right)^{-1} \mathbf{S}^\mathrm{F}_\mathcal{SA},
    \label{eq1}
\end{equation}
where $\mathcal{A}=\mathcal{R}\cup\mathcal{T}$, and $\mathcal{S}$ denotes the set of port indices associated with the NDA antennas. If the accessible antennas are partitioned into transmitting and receiving ones, (\ref{eq1}) specializes to
\begin{equation}
    \mathbf{H} = \mathbf{S}_\mathcal{RT} = \mathbf{S}^\mathrm{F}_\mathcal{RT} + \mathbf{S}^\mathrm{F}_\mathcal{RS} \left( \left(\mathbf{S}^\mathrm{L}\right)^{-1} - \mathbf{S}^\mathrm{F}_\mathcal{SS}\right)^{-1} \mathbf{S}^\mathrm{F}_\mathcal{ST}.
    \label{eq2}
\end{equation}

Our problem statement can now be formalized as follows: \textit{By measuring $\mathbf{S}$ (or $\mathbf{H}$) for a series of known realizations of $\mathbf{S}^\mathrm{L}$ (using the known, tunable load network), is it possible to subsequently estimate an unknown realization of $\mathbf{S}^\mathrm{L}$ (corresponding to the DUT) by measuring the associated $\mathbf{S}$ (or $\mathbf{H}$)?} 
In other words, is it possible to achieve a sufficiently unambiguous characterization of the OTA fixture based on known realizations of $\mathbf{S}^\mathrm{L}$ to subsequently de-embed the OTA fixture from a measurement taken with an unknown realization of $\mathbf{S}^\mathrm{L}$?

Our problem statement does \textit{not} require a tethered connection of any sort between the accessible, remote side and the NDA DUT side. Thus, it constitutes truly wireless sensing. As seen in Fig.~\ref{Fig1}, there are no RF cables connecting the two sides. Importantly, there is neither a need for cables to control nor to power the switches. Indeed, the switches can be powered autonomously using batteries or by harvesting power from wireless or solar energy.\footnote{In the case of wireless powering from the accessible antennas, the coherent input wavefront for optimal focusing on an NDA antenna can be determined by phase-conjugating the corresponding column of $\mathbf{S}^\mathrm{F}_\mathcal{SA}$~\cite{sol2024optimal}.} Moreover, the switches can cycle through a pre-defined sequence of configurations, removing the need for control cables. An experimental realization of battery-powered individual loads that autonomously cycle through a sequence of pre-defined configurations was already reported in~\cite{monsalve2013multiport}. For our proof-of-concept experiments reported below, however, we use cables to power and control the switches; we defer the development of an untethered tunable load network meeting our requirements to future work.

Before closing this section, we emphasize that our problem statement tacitly assumes that the waves radiated and captured via the accessible antennas can only interact with the DUT via the DUT's $N_\mathrm{S}$ monomodal ports. This is naturally the case in many practically relevant scenarios, for instance, if the DUT is a circuit or if the DUT is a network of sensors that encode information about humidity, temperature, etc. into their scattering parameters. 
However, the situation would be more complicated if the DUT were a radiating structure such as an antenna array. 
Although the radiation modes of an antenna array can be described in terms of scattering parameters~\cite{hansen1989relationships}, they are not associated with lumped (i.e., electrically small) ports. Hence, any scenario featuring coupling of the DUT to the WPE via radiation modes of the DUT is not compatible with our problem statement.\footnote{There are scenarios in which an antenna array could be considered as the DUT within our scope, for instance, if the WPE is an enclosure and the antenna array considered as the DUT is placed outside of this enclosure.} For this reason, we excluded this case at the outset of this section. To recapitulate, within the scope of our present paper, the DUT does not radiatively couple to the WPE that mediates the OTA coupling between the accessible antennas and the NDA antennas.

\section{Wireless Multi-Port Sensing Method}
\label{sec_method}

\subsection{Overview}

We proceed in three steps. \textit{First}, we characterize the OTA fixture based on the Virtual VNA technique, leveraging measurements of $\mathbf{S}$ or $\mathbf{H}$ corresponding to known realizations of $\mathbf{S}^\mathrm{L}$ (Fig.~\ref{Fig1}b). \textit{Second}, we connect the DUT to the NDA antennas and measure the corresponding $\mathbf{S}$ or $\mathbf{H}$ (Fig.~\ref{Fig1}c). \textit{Third}, we de-embed the OTA fixture from the measurement of $\mathbf{S}$ or $\mathbf{H}$ with the DUT to extract the DUT's scattering parameters. 

\subsection{Details}

\subsubsection{Step 1: OTA Fixture Characterization} Our goal in Step 1 is to determine a sufficiently unambiguous estimate of $\mathbf{S}^\mathrm{F}$ for the purposes of wireless multi-port sensing. The challenge lies in the fact that we can only inject and receive waves via a fixed subset of the OTA fixture's ports (specifically, those ports associated with the accessible antennas). However, we can terminate the remaining ports of the OTA fixture (those associated with NDA antennas) with different realizations of a known, tunable load network. The challenge of characterizing the OTA fixture under these constraints, summarized in Fig.~\ref{Fig1}b, almost exactly maps into the Virtual VNA setup in [Fig. 1a,~\cite{del2024virtual2p0}]. 
The only difference is that the setup in Fig.~\ref{Fig1}b does \textit{not} offer the possibility to connect one accessible and one NDA port via a coupled load. Indeed, such a coupled load is not feasible in the wireless sensing context of the present paper. Fortunately, such a coupled load is also not necessary here. Its absence merely results in an operationally insignificant ambiguity about the sign of $\mathbf{S}^\mathrm{F}_\mathcal{AS}=(\mathbf{S}^\mathrm{F}_\mathcal{SA})^\top$. By inspection, we see that if both $\mathbf{S}^\mathrm{F}_\mathcal{AS}$ and $\mathbf{S}^\mathrm{F}_\mathcal{SA}$ were multiplied by a minus sign in (\ref{eq1}), then these two minus signs would cancel out, implying that the ambiguity has no measurable effect in our problem statement. This is what we mean by ``sufficiently unambiguous'', i.e., we can tolerate this sign ambiguity here. However, all other requirements of the Virtual VNA technique are generally indispensable for wireless multi-port sensing; we consider the consequences of partially not satisfying them in our benchmarks detailed in Sec.~\ref{subsec_benchmarks_overview}.

As detailed in~\cite{del2024virtual,del2024virtual2p0}, there exist closed-form and gradient-descent variants of the Virtual VNA technique, requiring different sequences of configurations of the tunable load network. The gradient-descent approach tends to be more robust against noise~\cite{tapie2025scalable}, because it involves larger changes of the measurable $\mathbf{S}$ and it is flexible regarding the number of measurements.
Here, we use a \textit{novel} gradient-descent approach. In~\cite{del2024virtual,del2024virtual2p0,tapie2025scalable}, the gradient descent was applied to measurements obtained with random configurations of the tunable load network involving \textit{only} individual loads. Then, a few measurements for specific load network configurations involving the 2PLNs were used in~\cite{del2024virtual2p0,tapie2025scalable} to lift the remaining sign ambiguities in closed form. In contrast, here we do everything at once: We apply gradient descent to measurements obtained with random configurations of the tunable load network involving \textit{both} individual loads and 2PLNs. We ensure that each individual load and each 2PLN is used at least once. As evidenced by our benchmarking against simplified hardware in Sec.~\ref{sec_experimental_validation}, this novel gradient-descent procedure relaxes the requirements for the tunable load network for $N_\mathrm{S}>1$. Indeed, three distinct terminations of an NDA antenna do \textit{not} require the availability of three distinct individual loads when there are additional 2PLN terminations. This important insight had been overlooked in the Virtual VNA literature~\cite{del2024virtual,del2024virtual2p0,tapie2025scalable}. 

Our novel Virtual VNA gradient-descent procedure yields an estimate of $\mathbf{S}^\mathrm{F}$ (up to the sign ambiguity on $\mathbf{S}^\mathrm{F}_\mathcal{AS}=(\mathbf{S}^\mathrm{F}_\mathcal{SA})^\top$ that is operationally irrelevant here).
If we are constrained to measuring $\mathbf{H}$ instead of $\mathbf{S}$, we proceed analogously after replacing $\mathbf{S}^\mathrm{F}_\mathcal{AA}$, $\mathbf{S}^\mathrm{F}_\mathcal{AS}$ and $\mathbf{S}^\mathrm{F}_\mathcal{SA}$ with $\mathbf{S}^\mathrm{F}_\mathcal{RT}$, $\mathbf{S}^\mathrm{F}_\mathcal{RS}$ and $\mathbf{S}^\mathrm{F}_\mathcal{ST}$, respectively.

\subsubsection{Step 2: Measurement of DUT with OTA Fixture} In Step 2, we connect the NDA antennas to the unknown DUT and measure the corresponding $\mathbf{S}$ or $\mathbf{H}$.

\subsubsection{Step 3: De-Embedding the OTA Fixture} Our goal in Step 3 is to de-embed the OTA fixture from the measurement taken in Step 2, leveraging our characterization of the OTA fixture in Step 1, in order to obtain an estimate of the sought-after $\mathbf{S}^\mathrm{D}$. Given $\mathbf{S}^\mathrm{F}$ and $\mathbf{S}$ or $\mathbf{H}$, $\mathbf{S}^\mathrm{L}=\mathbf{S}^\mathrm{D}$ is the only unknown in (\ref{eq1}) or (\ref{eq2}), respectively. We opt for a gradient-descent-based estimation of $\mathbf{S}^\mathrm{L}$. We define the cost $\mathcal{C}$ to be minimized as the normalized sum of the element-wise absolute difference between measurement (MEAS) and prediction (PRED) of the scattering at the accessible antennas. Prediction refers to the value of $\mathbf{S}$ or $\mathbf{H}$ obtained upon injecting $\mathbf{S}^\mathrm{F}$ and our current estimate of $\mathbf{S}^\mathrm{L}$ into (\ref{eq1}) or (\ref{eq2}), respectively. 
Thus, if we measure $\mathbf{S}$,
\begin{equation}
\mathcal{C}
= \frac{
  \sum_{i,j}
    \bigl|\bigl(\mathbf{S}^{\mathrm{MEAS}} - \mathbf{S}^\mathrm{F}_\mathcal{AA}\bigr)_{ij}
          - \bigl(\mathbf{S}^{\mathrm{PRED}} - \mathbf{S}^\mathrm{F}_\mathcal{AA}\bigr)_{ij}\bigr|
}{
  \sum_{i,j}
    \bigl|\bigl(\mathbf{S}^{\mathrm{MEAS}} - \mathbf{S}^\mathrm{F}_\mathcal{AA}\bigr)_{ij}\bigr|
},
\end{equation}
 and if we measure $\mathbf{H}$,
\begin{equation}
\mathcal{C}
= \frac{
  \sum_{i,j}
    \bigl|\bigl(\mathbf{H}^{\mathrm{MEAS}} - \mathbf{S}^\mathrm{F}_\mathcal{RT}\bigr)_{ij}
          - \bigl(\mathbf{H}^{\mathrm{PRED}} - \mathbf{S}^\mathrm{F}_\mathcal{RT}\bigr)_{ij}\bigr|
}{
  \sum_{i,j}
    \bigl|\bigl(\mathbf{H}^{\mathrm{MEAS}} - \mathbf{S}^\mathrm{F}_\mathcal{RT}\bigr)_{ij}\bigr|
}.
\end{equation}
The only constraint imposed on $\mathbf{S}^\mathrm{L}$ is reciprocity.
The subtraction in these definitions of $\mathcal{C}$ removes the contributions of paths that never encounter the DUT and hence cannot help us with remotely sensing $\mathbf{S}^\mathrm{D}$.
The outcome of the gradient-descent procedure is our estimate of $\mathbf{S}^\mathrm{D}$. Explorations of alternative definitions of $\mathcal{C}$ or altogether alternative de-embedding techniques are deferred to future work.

\subsection{Benchmarks}
\label{subsec_benchmarks_overview}

We define the following benchmarks that arise from simplifying the tunable load network or the system model. 

\subsubsection{Simplifications of the Tunable Load Network}

The tunable load network in Fig.~\ref{Fig1} can terminate each NDA antenna with three distinct, individual loads or with coupled loads connecting it to a neighboring NDA antenna. 

If $N_\mathrm{S}=1$, there is only one possible hardware simplification benchmark:

\textit{A}: Each NDA antenna can be terminated with two (rather than three) distinct, individual loads.

If $N_\mathrm{S}>1$, there are three possible hardware simplification benchmarks:

\textit{A1}: Each NDA antenna can be terminated with three distinct, individual loads but there are no coupled loads.

\textit{A2}: Each NDA antenna can be terminated with two (rather than three) distinct, individual loads or with coupled loads connecting it to a neighboring NDA antenna.

\textit{A3}: Each NDA antenna can be terminated with two (rather than three) distinct, individual loads and there are no coupled loads.

\begin{figure*}[h]
    \centering
    \includegraphics[width=\textwidth]{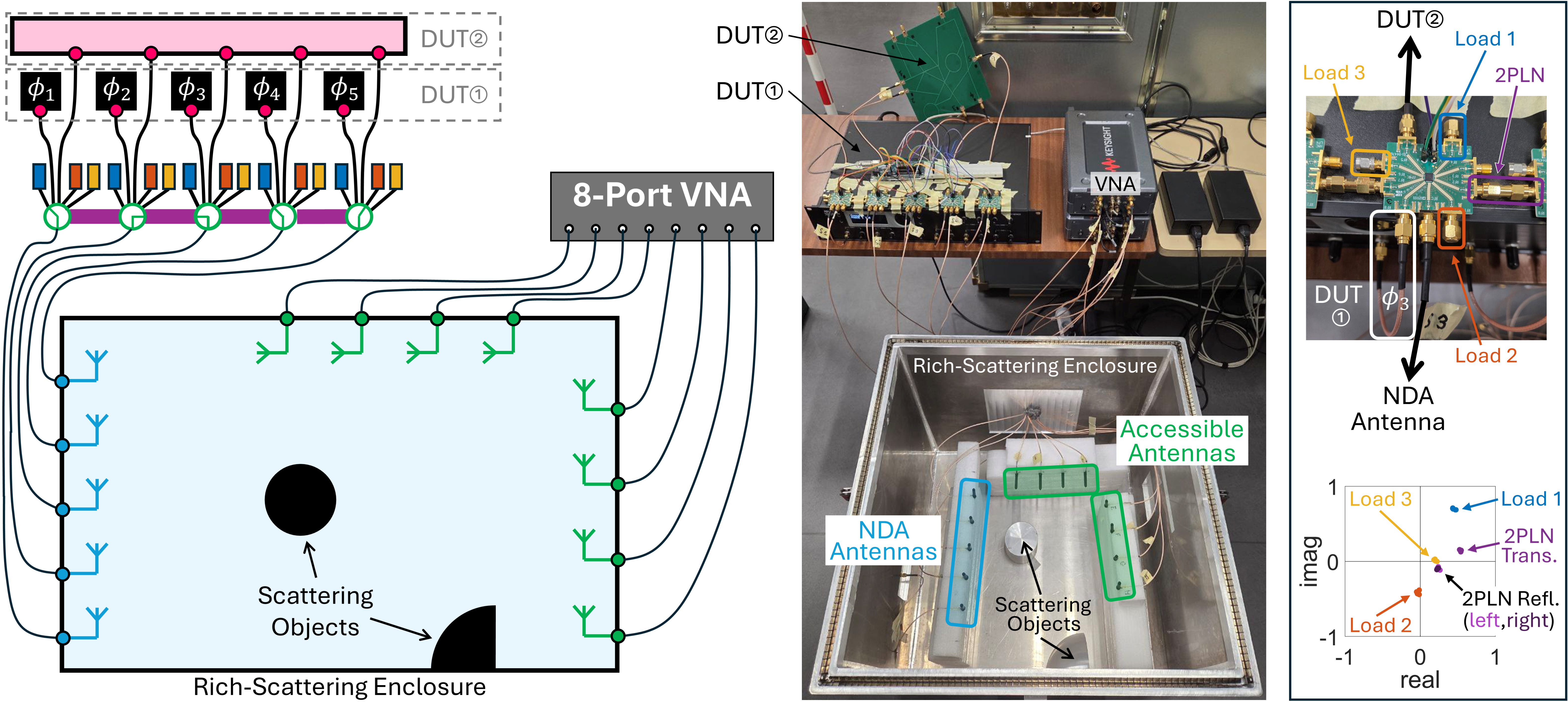}
    \caption{Schematic drawing and photographic images of the experimental setup (top cover removed to show interior). The scattering parameters of the tunable load network measured at 2.45~GHz are shown on the right; the scattering parameters of the two DUTs are shown as ground truth in Fig.~\ref{Fig3} and Fig.~\ref{Fig4}, respectively.}
    \label{Fig2}
\end{figure*}

\subsubsection{Simplified System Model}
\label{subsec_simplified_system_model}

For benchmark \textit{B}, we assume that all entries of $\mathbf{S}^\mathrm{F}_\mathcal{SS}$ vanish, implying zero mutual coupling between the NDA antennas \textit{and} perfect matching of the NDA antennas. This assumption yields an affine mapping from $\mathbf{S}^\mathrm{L}$ to $\mathbf{S}$ or $\mathbf{H}$:
\begin{subequations}
\label{eq3}
\begin{equation}
    \mathbf{S}^\mathrm{(1)} = \mathbf{S}^\mathrm{F}_\mathcal{AA} + \mathbf{S}^\mathrm{F}_\mathcal{AS} \mathbf{S}^\mathrm{L} \mathbf{S}^\mathrm{F}_\mathcal{SA},
    \label{eq3a}
\end{equation}
\begin{equation}
    \mathbf{H}^\mathrm{(1)}  = \mathbf{S}^\mathrm{(1)}_\mathcal{RT} = \mathbf{S}^\mathrm{F}_\mathcal{RT} + \mathbf{S}^\mathrm{F}_\mathcal{RS} \mathbf{S}^\mathrm{L} \mathbf{S}^\mathrm{F}_\mathcal{ST},
    \label{eq3b}
\end{equation}
\end{subequations}
where we use the superscript (1) to emphasize the approximative nature of (\ref{eq3}). In principle, assuming zero mutual coupling and perfect matching can introduce major inaccuracies. Nonetheless, reconfigurable intelligent surfaces (RISs) optimized under these assumptions performed almost as well as those optimized with the full, physics-compliant model (\ref{eq2}) in recent experiments~\cite{tapie2025beyond,del2025experimental}. However, here we tackle a sensing problem which may be more sensitive to the inaccuracies of the simplified system model.

Even though there is no mutual coupling between NDA antennas in the special case of $N_\mathrm{S}=1$, \textit{B} still implies an assumption of perfect matching. This assumption was made in a scenario with $N_\mathrm{S} = 1$ and $N_\mathrm{A}=2$ in~\cite{vena2024backscatter}. In this scenario, (\ref{eq2}) can be recast as $h = a+b(\frac{1}{c}-d)^{-1}$, where $h\in\mathbb{C}$ denotes $\mathbf{H}$, $a\in\mathbb{C}$ denotes $\mathbf{S}^\mathrm{F}_\mathcal{RT}$,  $b\in\mathbb{C}$ denotes $\mathbf{S}^\mathrm{F}_\mathcal{RS}\mathbf{S}^\mathrm{F}_\mathcal{ST}$, $c\in\mathbb{C}$ denotes $\mathbf{S}^\mathrm{L}$, and $d\in\mathbb{C}$ denotes $\mathbf{S}^\mathrm{F}_\mathcal{SS}$. Assuming that the NDA antenna is perfectly matched, i.e., $d=0$, $h$ simplifies to $h^\mathrm{(1)}=a+bc$, and the two unknown parameters ($a$ and $b$) characterizing the OTA fixture can be estimated via linear regression based on measurements of $h$ for two distinct, known values of $c$. This is precisely the approach taken in~\cite{vena2024backscatter}, which we have hereby rigorously related to multi-port network theory. However, if $d\neq 0$, the OTA fixture is characterized by three unknown parameters ($a$, $b$, and $d$) whose accurate estimation requires measurements of $h$ for three mutually distinct values of $c$. Moreover, if $d \neq 0$, the mapping from $c$ to $h$ is \textit{not} affine which prevents a parameter estimation  via linear regression.\footnote{Truncating the geometric series $h = a+b(\frac{1}{c}-d)^{-1} = a+bc\sum_{k=0}^\infty(dc)^k$ (whose convergence is guaranteed by the passivity of all considered scattering systems) after the first term featuring $d$, we find $h^\mathrm{(2)} = a + bc+bdc^2$. Based on this simplified model, the three unknown parameters characterizing the OTA fixture can be estimated via quadratic regression given measurements of $h$ for three distinct, known values of $c$. The $k$th term of the geometric series can be interpreted as a $k$-bounce path~\cite{del2025physics}. With $N_\mathrm{S}=1$, these bounces only occur at the interface between the OTA fixture and the DUT.}

\section{Experimental Validation}
\label{sec_experimental_validation}

In this section, we present an experimental proof-of-concept validation of our wireless multi-port sensing method. \textit{First}, we describe our experimental setup. \textit{Second}, we report our experimental results for 1-port DUTs (i.e., remote single-port load-impedance sensing). \textit{Third}, we report our experimental results for 5-port DUTs (i.e., remote multi-port load-impedance-matrix sensing).

\subsection{Experimental Setup}
\label{subsec_exp_setup}

We conduct our experiments at 2.45~GHz. Our setup comprises 8 accessible and 5 NDA antennas (ANT-24G-HL90-SMA). All NDA antennas have the same orientation; they are spaced by roughly half the operating wavelength along a line. The accessible antennas are split into two groups of four-element linear arrays with half-wavelength spacing; the antennas in these arrays are oriented perpendicular to each other to minimize the direct line-of-sight channel. 
As seen in Fig.~\ref{Fig2}, the $5+8=13$ antennas are placed inside a reverberation chamber (59~$\times$~60~$\times$~58~$\mathrm{cm}^3$) which yields a rich-scattering environment that precludes a priori channel knowledge. 

Each of the 5 NDA antennas is connected via a coaxial cable to the input port of a GaAs MMIC SP8T switch (HMC321ALP4E, Analog Devices). As shown in Fig.~\ref{Fig2}, the switch can connect its input port to one of seven output ports (there is also an  unused eighth output port). 
Three of these output ports are terminated by standard SMA loads (open circuit, short circuit, and matched load), and two output ports are connected via SMA connectors to neighboring switches. The corresponding measured scattering characteristics are displayed in the inset in Fig.~\ref{Fig2}; although not required, they are nearly identical for the five switches. None of the three realized individual loads emulates a calibration standard (because of the wave propagation in the cables and switches). In addition, each switch has one output port connected via a coaxial cable to a distinct channel of a reconfigurable phase shifter (API Weinschel, Inc. 8421-A1-08-FS) with 64 states. Furthermore, each switch has one output port connect via a coaxial cable to a distinct port of a complex, unknown 5-port circuit.

\begin{figure*}
    \centering
    \includegraphics[width=0.935\textwidth]{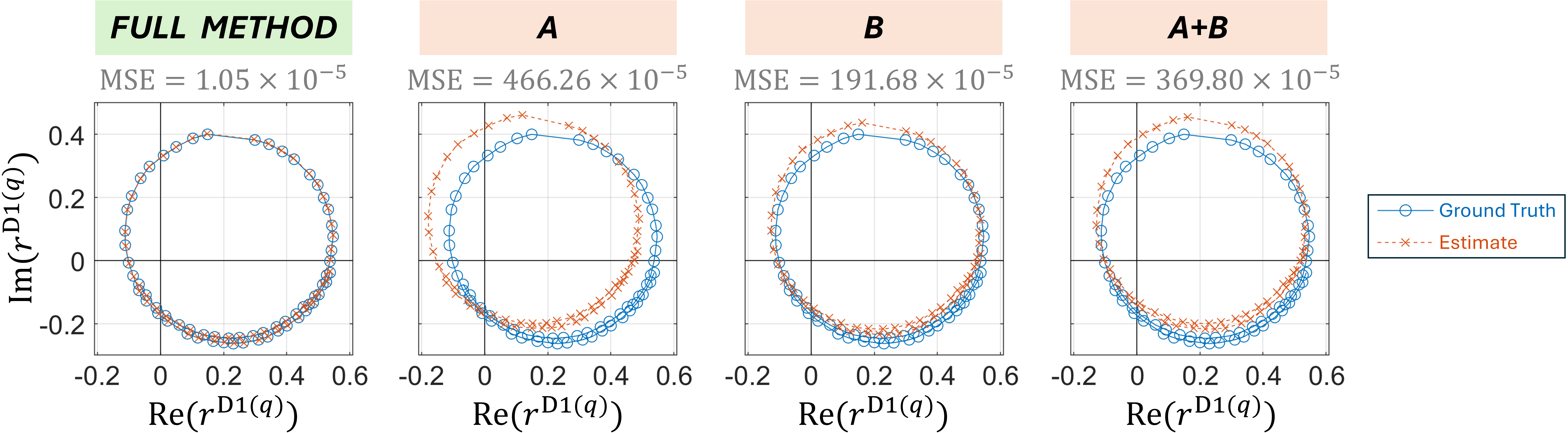}
    \caption{Selected experimental results for remote load-impedance sensing ($N_\mathrm{S}=1$) based on measurements of $\mathbf{H}$ with $N_\mathrm{A}=8$ and $n=30$. We visually contrast our estimate (red) of the 1-port DUT's reflection coefficient $r^{\mathrm{D1(}q\mathrm{)}}$ with the ground truth (blue), where $\mathrm{D1(}q\mathrm{)}$ denotes the $q$th state (out of 64) of DUT\textcircled{1}. Corresponding MSE values are indicated. Our full method is shown in the left panel, and three benchmarks (see definitions in Sec.~\ref{subsec_benchmarks_overview}) are shown in the remaining panels.}
    \label{Fig3}
\end{figure*}

This versatile setup satisfies the criteria for the tunable load network stated in Sec.~\ref{sec_problem_statement} and enables tests with different DUTs. 
Our DUT\textcircled{1} is the ensemble of 5 phase shifters, constituting a collection of five 6-bit programmable single-port loads without transmission between each other. Our DUT\textcircled{2} is the static 5-port circuit.  DUT\textcircled{2} is more generic than DUT\textcircled{1} in the sense that there is generally some transmission between its five ports.
Based on DUT\textcircled{1} we can also consider a 1-port DUT scenario by fixing the configuration of four switches to a reference termination (e.g., individual load 1 realized based on the open-circuit load) throughout the experiment. In that case, the four associated NDA antennas merely constitute environmental scattering objects in our rich-scattering WPE.

We use an eight-port VNA (two cascaded Keysight P5024B 4-port VNAs) to measure the $8\times 8$ scattering matrix at the ports of the eight accessible antennas. During data analysis, we can decide to only use certain entries of this measured scattering matrix to emulate a restriction to measuring a transmission matrix and/or to a lower number of accessible antennas.
We conduct our experiments under well-controlled conditions: The signal-to-noise ratio (SNR) is 63.1~dB (estimated based on repeated measurements in quick succession of $\mathbf{S}$ with all five switches in a reference configuration (individual load 1)) and the stability is 59.1~dB (estimated like the SNR, but based on measurements taken intermittently over the course of our experiment).

\subsection{Experimental Results for 1-Port DUTs}
\label{subsec_exp_1port}

We begin by presenting experimental results for remote load-impedance sensing for a 1-port DUT, meaning that four of the five switches are fixed to their reference state (individual load 1) throughout. For simplicity, we refer to $\mathbf{S}^\mathrm{D}$ as $r^\mathrm{D}$ in this case with $N_\mathrm{S}=1$. Moreover, the known, tunable individual load has only three distinct configurations for $N_\mathrm{S}=1$. There are three benchmarks: \textit{A}: using only two instead of three known individual loads; \textit{B}: using the simplified system model; \textit{A+B}: combining both simplifications. To quantify the accuracy of our estimate of $r^\mathrm{D}$, we report the mean-squared error (MSE) averaged over all 64 possible states of the 1-port DUT. 

A set of representative experimental results is displayed in Fig.~\ref{Fig3} for the case of measuring $\mathbf{H}$ with $N_\mathrm{A}=8$ and $n=30$. The results obtained with our full method shown in the left panel of Fig.~\ref{Fig3} are flawless upon visual inspection, and the corresponding MSE is very small. In contrast, the three simplification benchmarks in the remaining three panels of Fig.~\ref{Fig3} yield two orders of magnitude larger MSEs. Quantitative comparisons between these three benchmark MSEs are of limited value because of the specificity to our experiment, i.e., the dependence on the characteristics of the two available known individual loads and the NDA antenna's reflection coefficient in our WPE. According to our full method's estimate of $\mathbf{S}^\mathrm{F}_\mathcal{SS}$, the NDA antenna has a return loss of $-15.9$~dB in our WPE. It is hence reasonably well matched which likely limits the error of using the simplified system model. Qualitatively, the estimates of $r^\mathrm{D}$ clearly deviate from the ground truth for the three benchmarks in Fig.~\ref{Fig3}. Interestingly, however, certain topological relations between the 64 values of $r^\mathrm{D}$ appear to be recovered nonetheless (e.g., the circle shape, etc.). Although these estimates cannot rigorously be mapped onto the ground truth via shifts, rotations and tilts, maybe in certain applications the topological resemblance could be sufficient.

\begin{table}[b]
  \caption{MSE ($\times10^5$) for 1-port DUT ($N_\mathrm{S}=1$), using our full method with $n=30$.}
  \centering
  \begin{tabular}{|c||c|c|c|c|}
    \hline
    $N_\mathrm{A}$ & 8      & 6       & 4       & 2       \\
    \hline\hline
    $\mathbf{S}$     & 1.99 & 1.40 & 2.30 & 4.65 \\
    \hline
    $\mathbf{H}$     & 1.05 & 1.10 & 2.37 & 4.14 \\
    \hline
  \end{tabular}
  \label{Table1}
\end{table}

\begin{figure*}
    \centering
    \includegraphics[width=\textwidth]{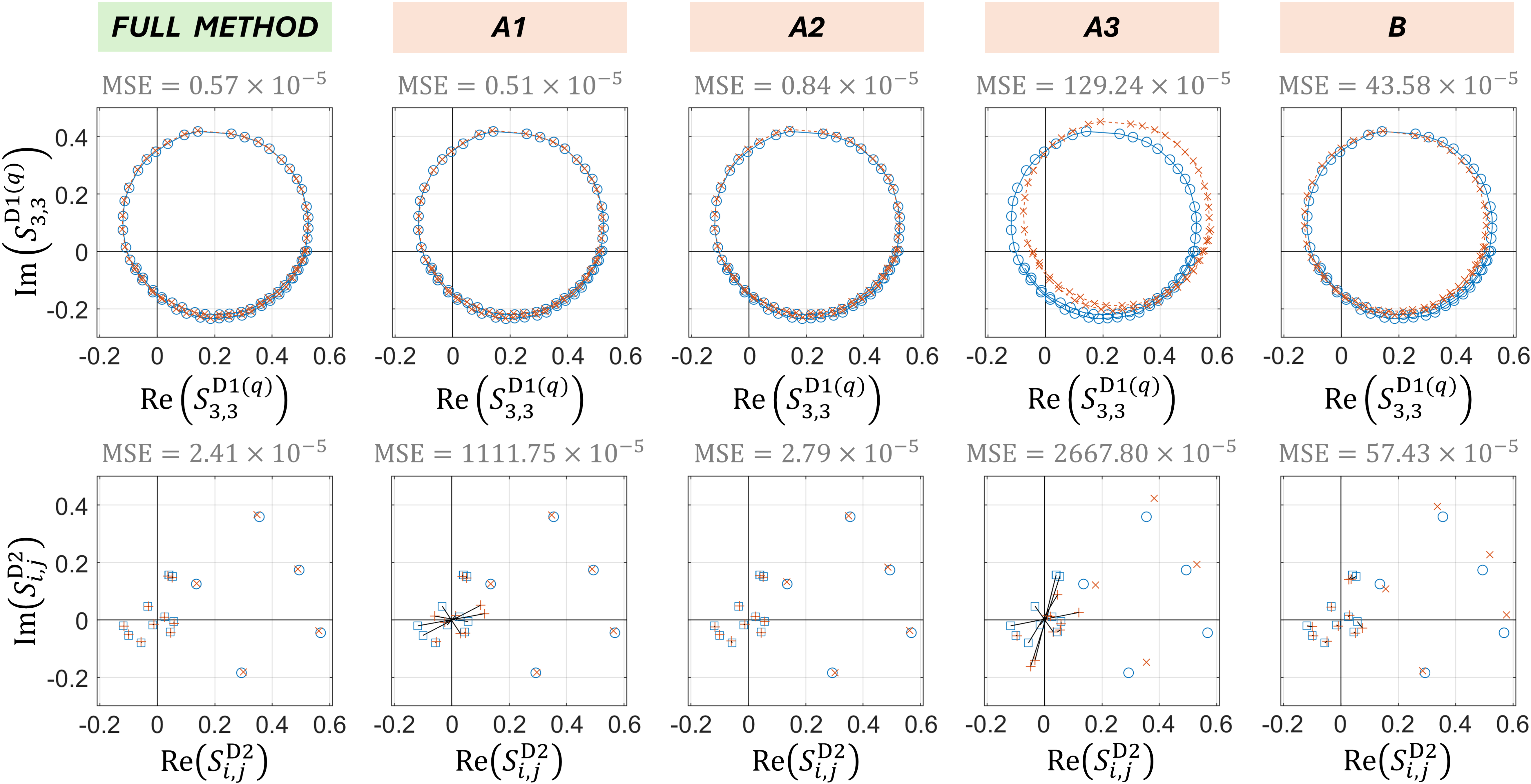}
    \caption{Selected experimental results for remote load-impedance-matrix sensing ($N_\mathrm{S}=5$) based on measurements of $\mathbf{S}$ with $N_\mathrm{A}=8$ and $n=300$. The first row considers the third diagonal entry of the scattering matrix of DUT\textcircled{1} in its $q$th state (out of 64), comparing our estimate (red cross) with the ground truth (blue circle).
    The second row considers all entries of the scattering matrix of DUT\textcircled{2}, comparing our estimate (red cross for diagonal entries, red plus for off-diagonal entries) with the ground truth (blue circle for diagonal entries, blue square for off-diagonal entries); black lines connect our estimates to the corresponding ground truths.
    Corresponding MSE values are indicated. Our full method is shown in the left panel, and four benchmarks (see definitions in Sec.~\ref{subsec_benchmarks_overview}) are shown in the remaining panels.}
    \label{Fig4}
\end{figure*}

We also examined the influence of the choice to measure $\mathbf{H}$ and the values of $N_\mathrm{A}$ and $n$. The value of $n$ had very little impact on our results because as soon as it exceeds the number of possible known individual load configurations it can be interpreted as merely constituting a means of averaging to effectively reduce the (already very low) measurement noise. The dependence of the achieved accuracy on measuring $\mathbf{H}$ vs $\mathbf{S}$ and the value of $N_\mathrm{A}$ is summarized in Table~\ref{Table1} for our full method with $n=30$. All MSEs are very small and of the same order of magnitude as the one seen in the left panel in Fig.~\ref{Fig3}. The MSE increases as $N_\mathrm{A}$ is reduced, which makes sense because the scattering measurements provide less information the fewer accessible antennas are used. Surprisingly, the results achieved upon measuring $\mathbf{H}$ are \textit{slightly} better than upon measuring $\mathbf{S}$. This is unexpected at first sight because $\mathbf{H}$ only offers a subset of the information contained in $\mathbf{S}$. 
Specifically, the reflection matrices of the transmitting and receiving antenna arrays ($\mathbf{S}_\mathcal{TT}$ and $\mathbf{S}_\mathcal{RR}$) are included in $\mathbf{S}$ but not in $\mathbf{H}$.

We hypothesize that a possible explanation may relate to different sensitivities of the blocks of $\mathbf{S}$ to $\mathbf{S}^\mathrm{D}$.
Because our definition of $\mathcal{C}$ in Step 3 uses $\mathbf{S}-\mathbf{S}^\mathrm{F}_\mathcal{AA}$ or $\mathbf{H}-\mathbf{S}^\mathrm{F}_\mathcal{RT}$, it only considers contributions from paths that encountered the DUT. However, given the orthogonal alignment of the transmitting and receiving arrays, it is likely that the dominant paths involved in $\mathbf{H}-\mathbf{S}^\mathrm{F}_\mathcal{RT}$ encountered the DUT many more times than the dominant paths involved in $\mathbf{S}_\mathcal{RR}-\mathbf{S}^\mathrm{F}_\mathcal{RR}$ and $\mathbf{S}_\mathcal{TT}-\mathbf{S}^\mathrm{F}_\mathcal{TT}$. A path's sensitivity to the DUT's scattering properties directly depends on how many times the DUT was encountered~\cite{del2021deeply,prod2024mutual,prod2025benefits}. Future work can explore how to refine the definition of $\mathcal{C}$ for Step 3 to optimally benefit from the additional information contained in $\mathbf{S}_\mathcal{TT}$ and $\mathbf{S}_\mathcal{RR}$.

\subsection{Experimental Results for 5-Port DUTs}
\label{subsec_exp_5port}

We now present experimental results for remote load-impedance-matrix sensing for reciprocal 5-port DUTs. Taking the reciprocity constraint into account, there are 15 independent, complex-valued entries of $\mathbf{S}^\mathrm{D}$ to be estimated.  The parameter estimation is well posed only if the number of independent, complex-valued parameters to be estimated equals or exceeds the number of independent, complex-valued measured parameters. The latter is summarized for various measurement conditions in Table~\ref{TableA}. Only measurements of $\mathbf{S}$ with $N_\mathrm{A}=8$ or $N_\mathrm{A}=6$ clearly exceed the number of sought-after parameters. Measurements with $\mathbf{H}$ and $N_\mathrm{A}=8$ barely exceed the number of sought-after parameters, implying a high vulnerability to inaccuracies and noise. In our experiments, we do not succeed to estimate  $\mathbf{S}^\mathrm{D}$ with $\mathbf{H}$ and $N_\mathrm{A}=8$; the achieved MSE is four orders of magnitude above the one obtained with $\mathbf{S}$ and $N_\mathrm{A}=8$ or $N_\mathrm{A}=6$. Thus, we only consider measurements of $\mathbf{S}$ in the remainder of this subsection.

\begin{table}[ht]
  \caption{Number of independent, complex-valued entries in $\mathbf{S}$ and $\mathbf{H}$.}
  \centering
  \begin{tabular}{|c||c|c|c|c|}
    \hline
    $N_\mathrm{A}$ & 8      & 6       & 4       & 2       \\
    \hline\hline
    $\mathbf{S}$     & 36 & 21 & 10 & 3 \\
    \hline
    $\mathbf{H}$     & 16 & 9 & 4 & 1 \\
    \hline
  \end{tabular}
  \label{TableA}
\end{table}

A set of representative experimental results is displayed in Fig.~\ref{Fig4} for the case of measuring $\mathbf{S}$ with $N_\mathrm{A}=8$ and $n=300$. We treat both DUTs as completely unconstrained except for reciprocity, i.e., we do \textit{not} use a priori knowledge about the vanishing off-diagonal entries of $\mathbf{S}^\mathrm{D}$ for DUT\textcircled{1}, nor that all diagonal entries are nominally identical for DUT\textcircled{1}. 
Upon visual inspection of Fig.~\ref{Fig4}, we observe that our full method yields flawless results, as also testified by very low MSE values, of the same order of magnitude as in Fig.~\ref{Fig3}. 

Benchmark \textit{A1} (no 2PLNs) only yields flawless results for DUT\textcircled{1} whereas sign errors on some off-diagonal entries for DUT\textcircled{2} are seen. The lines connecting pairs of ground truth and estimate all run (almost directly) through the origin, confirming that the only significant errors are sign errors. This makes sense because the absence of 2PLNs causes operationally significant sign ambiguities on off-diagonal entries of $\mathbf{S}^\mathrm{F}$ that propagate into the off-diagonal entries of our estimate of $\mathbf{S}^\mathrm{D}$. If, however, only the diagonal entries of $\mathbf{S}^\mathrm{D}$ or only the magnitudes of the entries of $\mathbf{S}^\mathrm{D}$ are sought, then the hardware simplification underlying benchmark \textit{A1} is perfectly acceptable.

Benchmark \textit{A2} (only two known individual loads) surprisingly yields flawless results on a par with those of our full method. As mentioned earlier, taking the 2PLNs into account, \textit{A2} in fact still enables at least three distinct terminations of the NDA antennas. Existing Virtual VNA literature~\cite{del2024virtual2p0,tapie2025scalable} had not recognized this important relaxation of hardware requirements for $N_\mathrm{S}>1$ regarding the number of known individual loads. 

Benchmark \textit{A3} (only two known individual loads and no 2PLNs), in contrast, yields three to four orders of magnitude larger MSE values and the estimates are clearly not coinciding with the ground truths. The hardware simplifications underlying \textit{A3} do not yield a sufficiently unambiguous estimate of $\mathbf{S}^\mathrm{F}$, and these ambiguities propagate into the estimate of $\mathbf{S}^\mathrm{D}$. As in Fig.~\ref{Fig3}, we note that important topological relations are present in the estimates nonetheless.

Benchmark \textit{B} (assuming all entries of $\mathbf{S}_\mathcal{SS}$ vanish) also yields large MSE values and the estimates do not coincide with the ground truths. The MSEs are an order of magnitude smaller than for \textit{A3} which is likely setup-specific rather than a general conclusion. According to our estimate of $\mathbf{S}^F_\mathrm{SS}$ with our full method, the return loss of our five NDA antennas is between $-12.9$~dB and $-22.1$~dB, and the transmission coefficients between our five NDA antennas are between $-16.7$~dB and $-30.7$~dB. Thus, our NDA antennas are reasonably well matched and feature reasonably low mutual coupling, which likely explains the limited error due to the inaccuracies of the simplified model assumed by \textit{B}.

We also examined the influence of the choice of the values of $N_\mathrm{A}$ and $n$ for DUT\textcircled{1}. As seen in Table~\ref{Table2}, reducing $n$ from 300 via 100 to 30 results in slight increases of the MSE. Reducing $N_\mathrm{A}$ from 8 to 6 notably increases the MSE (although it remains of the same order of magnitude). Reducing $N_\mathrm{A}$ further results in four orders of magnitudes larger MSEs because the parameter estimation problem becomes ill-posed, as discussed at the beginning of this subsection.

\begin{table}[ht]
  \caption{MSE ($\times10^5$) for 5-port DUT~\textcircled{1} ($N_\mathrm{S}=5$), using our full method for measurements of $\mathbf{S}$.}
  \centering
  \begin{tabular}{|c||c|c|c|c|}
    \hline
       & $N_\mathrm{A}=8$      & $N_\mathrm{A}=6$       & $N_\mathrm{A}=4$       & $N_\mathrm{A}=2$       \\
    \hline\hline
    $n=300$ & 0.57      & 1.65       & 2527.84       & 4586.69       \\
    \hline
    $n=100$     & 0.70 & 1.78 & 2442.20 & 4597.12 \\
    \hline
    $n=30$     & 0.81 & 2.19 & 2400.54 & 4650.48 \\
    \hline
  \end{tabular}
  \label{Table2}
\end{table}

\section{Conclusions}
\label{sec_conclusions}

To summarize, we have introduced and experimentally validated the first method for remote wireless sensing of a multi-port DUT's scattering parameters. We make no specific assumptions about the utilized antennas or the WPE. We connect a set of antennas to the DUT that couple OTA to a set of antennas via which we can inject and/or receive waves.
Our approach consists in interpreting the WPE (comprising the antennas) as an OTA fixture between our VNA and the DUT. By terminating the antennas that can be connected to the DUT initially with a known tunable load network instead of the DUT, we achieve a sufficiently unambiguous characterization of this OTA fixture based on Virtual VNA techniques. Subsequently, we de-embed the OTA fixture from measurements with the DUT. 
Our proof-of-concept experiments confirmed, for both 1-port DUTs and 5-port DUTs, that our method unambiguously recovers the DUT's full scattering matrix  with high accuracy. We  systematically examined key parameters influencing our OTA fixture characterization. Moreover, we systematically benchmarked our full method against simplified versions in terms of the hardware constraints and/or the system model. This benchmarking revealed a thus-far overlooked possible relaxation of the Virtual VNA's requirements for the tunable load network.

Looking forward, on the one hand, it will be enticing to develop versions of the presented wireless multi-port sensing technique that are purely based on non-coherent detection, which would significantly alleviate the hardware complexity. Existing literature on the Virtual VNA suggests indeed that coherent detection is dispensable~\cite{del2024virtual,del2024virtual2p0,del2025virtual3p1}. 
On the other hand, it will be interesting to explore how many-port DUTs can be characterized OTA without requiring a large number of accessible antennas and VNA ports. This could be be achieved based on compressed-sensing principles by leveraging the tunable load network in Step 2 to provide measurement diversity. Specifically, Step 2 would then require multiple measurements for which different subsets of NDA antennas would be connected to the DUT while the remaining NDA antennas would be connected to the tunable load network.

\section*{Acknowledgment}
The author acknowledges stimulating discussions with A.~Vena.

\bibliographystyle{IEEEtran}

\begin{thebibliography}{10}
\providecommand{\url}[1]{#1}
\csname url@samestyle\endcsname
\providecommand{\newblock}{\relax}
\providecommand{\bibinfo}[2]{#2}
\providecommand{\BIBentrySTDinterwordspacing}{\spaceskip=0pt\relax}
\providecommand{\BIBentryALTinterwordstretchfactor}{4}
\providecommand{\BIBentryALTinterwordspacing}{\spaceskip=\fontdimen2\font plus
\BIBentryALTinterwordstretchfactor\fontdimen3\font minus
  \fontdimen4\font\relax}
\providecommand{\BIBforeignlanguage}[2]{{%
\expandafter\ifx\csname l@#1\endcsname\relax
\typeout{** WARNING: IEEEtran.bst: No hyphenation pattern has been}%
\typeout{** loaded for the language `#1'. Using the pattern for}%
\typeout{** the default language instead.}%
\else
\language=\csname l@#1\endcsname
\fi
#2}}
\providecommand{\BIBdecl}{\relax}
\BIBdecl

\bibitem{brooker2013lev}
G.~Brooker and J.~Gomez, ``{Lev Termen's Great Seal bug analyzed},'' \emph{IEEE
  Aerosp. Electron. Syst. Mag.}, vol.~28, no.~11, pp. 4--11, 2013.

\bibitem{Marrocco}
G.~Marrocco, ``Pervasive electromagnetics: sensing paradigms by passive {RFID}
  technology,'' \emph{IEEE Wirel. Commun.}, vol.~17, no.~6, pp. 10--17, 2010.

\bibitem{Marrocco_RFID_GRID}
G.~Marrocco, ``{RFID} grids: Part {I}—{E}lectromagnetic theory,'' \emph{IEEE
  Trans. Antennas Propag.}, vol.~59, no.~3, pp. 1019--1026, 2011.

\bibitem{Marrocco_RFID_GRID2}
S.~Caizzone and G.~Marrocco, ``{RFID} grids: Part {II}—{E}xperimentations,''
  \emph{IEEE Trans. Antennas Propag.}, vol.~59, no.~8, pp. 2896--2904, 2011.

\bibitem{marrocco2008multiport}
G.~Marrocco \emph{et~al.}, ``Multiport sensor {RFIDs} for
  wireless passive sensing of objects—{B}asic theory and early results,''
  \emph{IEEE Trans. Antennas Propag.}, vol.~56, no.~8, pp. 2691--2702, 2008.

\bibitem{mughal2023statistical}
A.~Mughal \emph{et~al.}, ``Statistical evaluation of the coupling effects between tags in a {UHF RFID}
  forward link,'' \emph{IEEE J. Radio Freq. Identif.}, vol.~7, pp. 257--266,
  2023.

\bibitem{nanni2024stackedRFID}
F.~M.~C. Nanni and G.~Marrocco, ``Experimental evaluation and upper-bounds of
  cross-sensitivity in stacked {RFID} sensors,'' \emph{IEEE J. Radio Freq.
  Identif.}, vol.~8, pp. 98--104, 2024.

\bibitem{barbot2025differential}
N.~Barbot \emph{et~al.}, ``Differential {RCS} of multi-port tag
  antenna with synchronous modulated backscatter,'' \emph{IEEE J. Radio Freq.
  Identif.}, 2025.

\bibitem{caizzone2011multi}
S.~Caizzone \emph{et~al.}, ``Multi-chip {RFID} antenna
  integrating shape-memory alloys for detection of thermal thresholds,''
  \emph{IEEE Trans. Antennas Propag.}, vol.~59, no.~7, pp. 2488--2494, 2011.

\bibitem{garbacz1964determination}
R.~Garbacz, ``Determination of antenna parameters by scattering cross-section
  measurements,'' \emph{Proc. Inst. Electr. Eng.}, vol. 111, no.~10, pp.
  1679--1686, 1964.

\bibitem{mayhan1994technique}
J.~T. Mayhan \emph{et~al.}, ``A technique for measuring
  antenna drive port impedance using backscatter data,'' \emph{IEEE Trans.
  Antennas Propag.}, vol.~42, no.~4, pp. 526--533, 1994.

\bibitem{pursula2008backscattering}
P.~Pursula \emph{et~al.}, ``Backscattering-based measurement
  of reactive antenna input impedance,'' \emph{IEEE Trans. Antennas Propag.},
  vol.~56, no.~2, pp. 469--474, 2008.

\bibitem{bories2010small}
S.~Bories \emph{et~al.}, ``Small antennas
  impedance and gain characterization using backscattering measurements,''
  \emph{Proc. EuCAP}, 2010.

\bibitem{van2020verification}
A.~J. Van Den~Biggelaar \emph{et~al.}, ``Verification of a contactless characterization method for
  millimeter-wave integrated antennas,'' \emph{IEEE Trans. Antennas Propag.},
  vol.~68, no.~5, pp. 3358--3365, 2020.

\bibitem{sahin2021noncontact}
S.~Sahin \emph{et~al.}, ``Noncontact characterization of antenna
  parameters in {mmW} and {THz} bands,'' \emph{IEEE Trans. Terahertz Sci.
  Technol.}, vol.~12, no.~1, pp. 42--52, 2021.

\bibitem{kruglov2023contactless}
D.~Kruglov \emph{et~al.}, ``Contactless measurement of a {D}-band on-chip antenna using an integrated
  reflective load switch,'' \emph{IEEE Antennas Wirel. Propag. Lett.}, vol.~23,
  no.~3, pp. 1075--1079, 2023.

\bibitem{del2024virtual}
P.~del Hougne, ``{Virtual VNA}: Minimal-ambiguity scattering matrix estimation
  with a fixed set of ``virtual'' load-tunable ports,'' \emph{IEEE Trans.
  Instrum. Meas.}, vol.~74, pp. 1--19, 2025.

\bibitem{wiesbeck1998wide}
W.~Wiesbeck and E.~Heidrich, ``Wide-band multiport antenna characterization by
  polarimetric {RCS} measurements,'' \emph{IEEE Trans. Antennas Propag.},
  vol.~46, no.~3, pp. 341--350, 1998.

\bibitem{monsalve2013multiport}
B.~Monsalve \emph{et~al.}, ``Multiport small integrated antenna
  impedance matrix measurement by backscattering modulation,'' \emph{IEEE
  Trans. Antennas Propag.}, vol.~61, no.~4, pp. 2034--2042, 2013.

\bibitem{shilinkov2024antenna}
I.~Shilinkov and R.~Maaskant, ``Antenna array measurements by a scalable
  backscatter modulation procedure,'' \emph{IEEE Antennas Wirel. Propag.
  Lett.}, vol.~23, no.~10, pp. 2989--2993, 2024.

\bibitem{denicke2012application}
E.~Denicke \emph{et~al.}, ``The application of multiport
  theory for {MIMO RFID} backscatter channel measurements,'' \emph{Proc. EuMC},
  pp. 522--525, 2012.

\bibitem{del2024virtual2p0}
P.~del Hougne, ``Virtual {VNA 2.0}: Ambiguity-free scattering matrix estimation
  by terminating not-directly-accessible ports with tunable and coupled
  loads,'' \emph{IEEE Trans. Antennas Propag.}, vol.~73, no.~7, pp. 4903--4908,
  2025.

\bibitem{tapie2025scalable}
J.~Tapie and P.~del Hougne, ``Scalable multiport antenna array characterization
  with {PCB}-realized tunable load network providing additional “virtual”
  {VNA} ports,'' \emph{IEEE Antennas Wirel. Propag. Lett.}, 2025.

\bibitem{bauer1974embedding}
R.~F. Bauer and P.~Penfield, ``De-embedding and unterminating,'' \emph{IEEE
  Trans. Microw. Theory Techn.}, vol.~22, no.~3, pp. 282--288, 1974.

\bibitem{davidovitz1995reconstruction}
M.~Davidovitz, ``Reconstruction of the {S}-matrix for a 3-port using
  measurements at only two ports,'' \emph{IEEE Microw. Guid. Wave Lett.},
  vol.~5, no.~10, pp. 349--350, 1995.

\bibitem{lu2000port}
H.-C. Lu and T.-H. Chu, ``Port reduction methods for scattering matrix
  measurement of an n-port network,'' \emph{IEEE Trans. Microw. Theory Techn.},
  vol.~48, no.~6, pp. 959--968, 2000.

\bibitem{lu2003multiport}
H.-C. Lu and T.-H. Chu, ``Multiport scattering matrix measurement using a reduced-port network
  analyzer,'' \emph{IEEE Trans. Microw. Theory Techn.}, vol.~51, no.~5, pp.
  1525--1533, 2003.

\bibitem{pfeiffer2005recursive}
U.~R. Pfeiffer and C.~Schuster, ``A recursive un-termination method for
  nondestructive in situ {S}-parameter measurement of hermetically encapsulated
  packages,'' \emph{IEEE Trans. Microw. Theory Techn.}, vol.~53, no.~6, pp.
  1845--1855, 2005.

\bibitem{pfeiffer2005equivalent}
U.~Pfeiffer and B.~Welch, ``Equivalent circuit model extraction of flip-chip
  ball interconnects based on direct probing techniques,'' \emph{IEEE Microw.
  Wirel. Compon. Lett.}, vol.~15, no.~9, pp. 594--596, 2005.

\bibitem{pfeiffer2005characterization}
U.~R. Pfeiffer and A.~Chandrasekhar, ``Characterization of flip-chip
  interconnects up to millimeter-wave frequencies based on a nondestructive in
  situ approach,'' \emph{IEEE Trans. Adv. Packag.}, vol.~28, no.~2, pp.
  160--167, 2005.

\bibitem{del2025virtual3p0}
P.~del Hougne, ``Virtual {VNA 3.0}: Unambiguous scattering matrix estimation
  for non-reciprocal systems by leveraging tunable and coupled loads,''
  \emph{arXiv:2503.07239}, 2025.

\bibitem{del2025virtual3p1}
P.~del Hougne, ``Virtual {VNA 3.1}: Non-coherent-detection-based non-reciprocal
  scattering matrix estimation leveraging a tunable load network,''
  \emph{arXiv:2504.11790}, 2025.

\bibitem{chen2012wireless}
H.-Y. Chen \emph{et~al.}, ``Wireless impedance measurement of {UHF RFID} tag chips,'' \emph{Proc. IMS},
  pp. 1--3, 2012.

\bibitem{bjorninen2011wireless}
T.~Bjorninen \emph{et~al.}, ``Wireless measurement of {RFID IC} impedance,'' \emph{IEEE
  Trans. Instrum. Meas.}, vol.~60, no.~9, pp. 3194--3206, 2011.

\bibitem{akbar2015rfid}
M.~Akbar \emph{et~al.}, ``{RFID} tag load
  impedance measurement using backscattered signal,'' \emph{IEEE Int. Symp.
  Antennas Propag. USNC/URSI Natl. Radio Sci. Meet.}, pp. 1762--1763, 2015.

\bibitem{skrobacz2024new}
K.~Skrobacz \emph{et~al.}, ``A new concept of determining the {RFID} chip impedance,'' \emph{IEEE Trans.
  Microw. Theory Tech.}, 2024.

\bibitem{chen2012coupling}
H.-Y. Chen \emph{et~al.}, ``Coupling passive sensors to {UHF RFID} tags,'' \emph{Proc. IEEE Radio
  Wirel. Symp.}, pp. 255--258, 2012.

\bibitem{vena2024backscatter}
A.~Vena \emph{et~al.}, ``Backscatter-based wireless sensing system for
  multi-channel complex impedance measurements,'' \emph{Proc. Int. Conf. Smart
  Sustain. Tech.}, pp. 1--4, 2024.

\bibitem{anderson_cascade_1966}
B.~D.~O. Anderson and R.~W. Newcomb, ``\BIBforeignlanguage{en}{Cascade
  connection for time-invariant n-port networks},''
  \emph{\BIBforeignlanguage{en}{Proc. Inst. Electr. Eng.}}, vol. 113, no.~6,
  pp. 970--974, Jun. 1966.

\bibitem{ha1981solid}
T.~T. Ha, \emph{{Solid-State Microwave Amplifier Design}}. Wiley-Interscience, 1981.

\bibitem{sol2024optimal}
J.~Sol \emph{et~al.}, ``Optimal blind focusing on perturbation-inducing targets
  in sub-unitary complex media,'' \emph{Laser Photonics Rev.}, p. 2400619,
  2024.

\bibitem{hansen1989relationships}
R.~C. Hansen, ``Relationships between antennas as scatterers and as
  radiators,'' \emph{Proc. IEEE}, vol.~77, no.~5, pp. 659--662, 1989.

\bibitem{tapie2025beyond}
J.~Tapie \emph{et~al.}, ``Beyond-diagonal {RIS} prototype and performance
  evaluation,'' \emph{arXiv:2505.13392}, 2025.

\bibitem{del2025experimental}
P.~del Hougne, ``Experimental multiport-network parameter estimation and
  optimization for multi-bit {RIS},'' \emph{arXiv:2507.02168}, 2025.

\bibitem{del2025physics}
P.~del Hougne, ``A physics-compliant diagonal representation for wireless channels
  parametrized by beyond-diagonal reconfigurable intelligent surfaces,''
  \emph{IEEE Trans. Wirel. Commun.}, 2025.

\bibitem{del2021deeply}
M.~del Hougne \emph{et~al.}, ``Deeply subwavelength localization with
  reverberation-coded aperture,'' \emph{Phys. Rev. Lett.}, vol. 127, p. 043903,
  Jul 2021.

\bibitem{prod2024mutual}
H.~Prod'homme and P.~del Hougne, ``Mutual coupling in dynamic metasurface
  antennas: Foe, but also friend,'' \emph{arXiv:2412.01002}, 2024.

\bibitem{prod2025benefits}
H.~Prod'homme \emph{et~al.}, ``Benefits of mutual coupling in dynamic
  metasurface antennas for optimizing wireless communications - theory and
  experimental validation,'' \emph{arXiv:2502.15565}, 2025.

\end{thebibliography}

\providecommand{\noopsort}[1]{}\providecommand{\singleletter}[1]{#1}%

\end{document}